\documentclass[aps,twocolumn,superscriptaddress,floatfix,nofootinbib]{revtex4-2}
\usepackage[utf8]{inputenc}
\usepackage{amssymb}
\usepackage[dvipsnames]{xcolor}
\usepackage{amsmath}
\usepackage{graphicx}
\usepackage{appendix}
\usepackage{algorithmic}
\usepackage{float}
\makeatletter
\let\newfloat\newfloat@ltx
\makeatother
\usepackage{algorithm}
\usepackage{placeins}
\usepackage{bm}
\usepackage{lipsum}

\begin{document}

\global\long\def\ket#1{\left|#1\right\rangle }%

\global\long\def\bra#1{\left\langle #1\right|}%

\global\long\def\linner#1#2{\left\langle \left.#1\,\right|#2\hspace{1.2pt}\right\rangle }%

\global\long\def\rinner#1#2{\left\langle \hspace{1.2pt}#1\left|\,#2\right.\right\rangle }%

\newcommand{\code}[1]{\texttt{#1}}

\title{Supply Chain Logistics with Quantum and Classical Annealing Algorithms}

\author{Sean J. Weinberg}
\affiliation{QC Ware Corp., Palo Alto, CA USA}
\author{Fabio Sanches}
\affiliation{QC Ware Corp., Palo Alto, CA USA}
\author{Takanori Ide}
\affiliation{
AISIN CORPORATION, Tokyo Research Center, Chiyoda-ku, Tokyo, Japan
}
\author{Kazumitzu Kamiya}
\affiliation{Aisin Technical Center of America, San Jose, CA  USA}
\author{Randall Correll}
\affiliation{QC Ware Corp., Palo Alto, CA USA}

\date{\today}

\begin{abstract}

Noisy intermediate-scale quantum (NISQ) hardware is almost universally incompatible with full-scale optimization problems of practical importance which can have many variables and unwieldy objective functions. As a consequence, there is a growing body of literature that tests quantum algorithms on miniaturized versions of problems that arise in an operations research setting. Rather than taking this approach, we investigate a problem of substantial commercial value, multi-truck vehicle routing for supply chain logistics, at the scale used by a corporation in their operations. Such a problem is too complex to be fully embedded on any near-term quantum hardware or simulator; we avoid confronting this challenge by taking a hybrid workflow approach: we iteratively assign routes for trucks by generating a new binary optimization problem instance one truck at a time. Each instance has $\sim 2500$ quadratic binary variables, putting it in a range that is feasible for NISQ quantum computing, especially quantum annealing hardware. We test our methods using simulated annealing and the D-Wave Hybrid solver as a place-holder in wait of quantum hardware developments. After feeding the vehicle routes suggested by these runs into a highly realistic classical supply chain simulation, we find excellent performance for the full supply chain. Our work gives a set of techniques that can be adopted in contexts beyond vehicle routing to apply NISQ devices in a hybrid fashion to large-scale problems of commercial interest.

\end{abstract}

\maketitle

\section{Introduction}

Quantum algorithms have the capacity to offer enormous
performance improvements over known classical algorithms
for solving important problems like integer factoring and quantum mechanical simulation \cite{shor1999polynomial, feynman2018simulating, lloyd1996universal}. However, despite extraordinary effort and investment,
the current state of quantum hardware remains too immature
for there to be practical computational value from any
quantum device that exists today.  Recently, calculations have been performed
with a quantum computer that outperform classical computing \cite{arute2019quantum},
but these calculations do not solve a problem of practical use.

Given these realities, much research has been devoted to heuristic
algorithms, often of a hybrid classical-quantum nature, that aim
to offer computational advantage even with problematic hardware \cite{farhi2014quantum, kadowaki1998quantum, farhi2000quantum, peruzzo2014variational, preskill2018quantum}.  
While such algorithms that are suitable for the noisy intermediate-scale quantum era (NISQ) \cite{preskill2018quantum} are not proven to offer complexity
advantages over classical methods, they have potential to become an oasis during the long road to the development of large-scale fault-tolerant quantum computers.

NISQ algorithms and rapid development of quantum hardware
has caught the attention of numerous industries that are in search
of solutions that can improve efficiency in their operations. 
To test the potential for NISQ algorithms in a given use case,
it is commonplace to begin with a complex problem and to then
identify a miniaturized version of the problem which can then
be attacked with a NISQ quantum algorithm. For example,
many problems that arise in operations research can be recast
as a quadratic unconstrained binary optimization (QUBO) problem.
Such problems can be downsized and then solved using the quantum
approximate optimization algorithm \cite{farhi2014quantum} (with simulated or real hardware) or by execution on quantum annealing
devices like those produced by D-Wave Systems \cite{johnson2011quantum}.

There is no question that such studies are of value, but there
are also major issues with them. Problem instances must be reduced to
very small toy models that barely reflect the true nature of practical
problems. Circuit model quantum algorithms cannot currently be tested, even on simulators, much beyond 30 binary variables. Quantum annealers can become cumbersome around a few hundred variables due to embedding challenges. Private companies may not be inclined to fund quantum computing studies when restricted to such small use cases and when the quality of solution and runtime is all but guaranteed to be easily outperformed by classical algorithms with readily available hardware.

In this work, we develop an alternative approach to downsizing problems
for commercial studies. We begin with a realistic vehicle routing
problem that arises in a company's operations. Rather than distilling
the mathematical problem and constructing a version with very few variables, we instead use a hybrid workflow approach: we iteratively
construct small QUBO instances that are of reasonable size for near-term  hardware with NISQ algorithms. Solutions to these small instances
do not solve the full and very complex vehicle routing problem,
but they do provide a route for a single truck. We use that route
to update the remaining unfulfilled demand and we then repeat
the procedure.  After obtaining solutions for all trucks, we input the routes into a highly realistic simulation of the flow of all trucks and carried boxes, taking into account various constraints that are difficult to include in QUBO instances. The final result is a viable solution to the full-scale routing problem which can be compared to solutions obtained through other means.

Using iterative approaches to build up a heuristic solution to a complex problem instance is certainly not a new idea. However, the construction
we give is carefully crafted to suit the needs of near-term quantum
computing hardware: the problem instances at each step are QUBO instances
with no more than a few thousand variables. These QUBO instances will be appropriate for circuit model quantum hardware with a few thousand noisy
qubits. In the more near term, D-Wave Systems has quantum annealing
hardware that can optimize QUBO instances with thousands of variables
already. However, our instances are just slightly too large for direct application
on D-Wave annealers due to the difficulty of embedding QUBO connectivity graphs.

The construction described in this work yields a viable quantum-classical
solution for a specific problem of substantial commercial value. 
Specifically, we study the problem of routing trucks in the supply
chain of Aisin Corporation, a Japanese automotive component manufacturer.
We emphasize, however, that the broad approach of deploying quantum
algorithms on small problem instances to build up a hybrid solution
to full-scale commercial applications is important well outside
of the context of supply chain optimization, and in fact such approaches
may be the first ways that quantum computing will be used outside of a research setting.

\subsection*{Outline}
In section \ref{sec:Supply-Chain-Optimization}, we explain 
the vehicle routing problem (VRP) that models that Aisin Corporation
supply chain. There are many ingredients to the routing problem,
so we gradually build it up by starting with a well-known vehicle
routing problem and adding various ingredients in a step-by-step
fashion. Section \ref{sec:Single-Truck-PUBO} describes the structure of binary optimization problem instances that we construct for each
truck in an iterative fashion. These are the optimization problems
that can be submitted to quantum hardware in a quantum implementation
of our methods. In section \ref{sec:Supply-Chain-Workflow}, we
describe the algorithm for iteratively constructing the optimization
problems from section \ref{sec:Single-Truck-PUBO}, and using their
solutions to update initial conditions related to the remaining unfulfilled demand. In subsection \ref{sec:truck-loop-execution}, we execute our workflow. As a
place-holder for full quantum solutions to the binary optimization
problem instances, we use simulated annealing and the ``D-Wave hybrid''
tool provided by D-Wave Systems. These solution methods offer insight
into how the workflow might behave when a direct quantum algorithmic
approach becomes feasible. Finally, in subsection \ref{sec:full-scale-sim}, we take input the route solutions from our simulated annealing and D-Wave hybrid runs into a full-scale supply chain simulation which tracks, in
detail, the locations of all boxes and trucks, ensuring that 
realistic and cumbersome constraints are satisfied. This allows us
to evaluate the overall performance of our methods by determining the
percentage of demand that is fulfilled given the number of trucks used.

\section{Vehicle Routing Problems\label{sec:Supply-Chain-Optimization}}

Although there are many aspects of supply chains that can be optimized,
we focus on vehicle routing. Mathematically, the wide variety of
combinatorial optimization problems go by the umbrella name of 
``vehicle routing problems'' (VRP) \cite{dantzig1959truck,toth2002vehicle}. The VRP variant studied in this paper is closely modeled after
the supply chain of Aisin Corporation, and is thus very
complex. It involves many trucks, nodes, and goods that must be carried
by trucks in ways that satisfy certain routing constraints. We explain
this vehicle routing problem by building it up, starting from a very
basic one.

\subsection{Basic Vehicle Routing Problem\label{subsec:Basic-Vehicle-Routing}}

The VRP discussed in this section is most similar to the split-delivery
capacitated vehicle routing problem, but since it's our base example
from which we will build up a more realistic supply chain model, we
refer to it informally as the ``basic VRP''. An instance of the
basic VRP is specified by:
\begin{enumerate}
\item A graph $G$ with $n + 1$ nodes. For convenience, the nodes are labelled
as $z_{0},z_{1},\ldots,z_{n}$
\item One special node $z_{0}$ selected from the graph which is called
the \emph{depot.}
\item An $n+1\times n+1$ matrix $T$ with nonnegative entries and $T_{ii}=0$
for all $i\in\{0,1,\ldots,n\}$ called the \emph{time matrix}.
\item A nonnegative number $d_{i}$ assigned to each node $z_{i}$ except
for the depot ($i\neq0$). These numbers are called \emph{initial
demands.}
\end{enumerate}
The nodes of the graph can be abstract, but in many cases they are
explicitly given as coordinates for locations that trucks might drive
to. The time matrix entry $T_{ij}$ is supposed to be the time it
take a truck to drive from node $i$ to node $j$. That is why we
require $T_{ii}=0$. The initial demand $d_{i}$ is supposed to be
the amount of material\footnote{``Amount of material'' is intentionally vague. In a practical application,
demand can be quantified by geometrical volume, by weight, or even
by monetary value. For our purposes, we will use geometrical volume
as the standard meaning for ``amount of material'' which makes it
sensible that trucks have a limited carrying capacity.} that must be carried by a truck from the node $z_{j}$ to the depot
node $z_{0}$.

A candidate solution to this VRP is given by a route: a list of integers
$\xi=\xi_{1},\xi_{2},\ldots,\xi_{k}$ where $k$ is some positive
integer (called the route length) and each $\xi_{j}$ is an element
of $\{0,1,\ldots,n\}$. We require that $\xi_{1}=0$ and $\xi_{k}=0$
so that the truck starts and ends at the depot. Given such a sequence,
there are two questions:
\begin{itemize}
\item What is the total driving time for $\xi$?
\item Is $\xi$ a demand-satisfying route?
\end{itemize}
The driving time for the route $\xi$ is computed in the obvious way:
\begin{equation}
\text{time}(\xi)=\sum_{t=1}^{k-1}T_{\xi_{t}\xi_{t+1}}.\label{eq:time_basic_vrp}
\end{equation}

The question of whether or not the route is demand-satisfying requires
some further elaboration. In short, a route is demand satisfying if
it will result in a truck carrying all of the initial demand to the
depot. The mathematical description of this concept is somewhat inelegant
due to the nature of picking up material from one location and bringing
it to another. The truck which starts at $\xi_{1}$ and follows the
route has a \emph{capacity} which we always take to be 1. This is
the amount of demand that the truck can store. The amount of demand
the truck is carrying at a given time is called \emph{on-board demand}
and the truck's on-board demand is initially zero, and we denote this
by $e_{1}=0$. (The index $1$ on $e_{1}$ refers to the fact that
it is the on-board demand after stopping at $\xi_{1}=0$ which is
the first time step.) Then, the truck drives to $\xi_{2}$ and picks
up demand from $\xi_{2}$. The amount picks up is largest allowed
amount:
\[
e_{2}=\min\left(1,d_{\xi_{2}}\right)=\min\left(1,e_{1}+d_{\xi_{2}}\right).
\]
Note that this is either the full demand at node $\xi_{2}$ or it
is a portion of that demand that fills up the truck to capacity. After
this step, we update the \emph{off-board} demand located at node $\xi_{2}$.
This means that we change $d_{\xi_{2}}$ by reducing it by the amount
that the truck took from $\xi_{2}$:
\[
d_{\xi_{2}}^{2}=d_{\xi_{2}}-\left(e_{2}-e_{1}\right).
\]
Explicitly writing $e_{2}-e_{1}$ even though $e_{1}=0$ is pedantic, but it helps clarify how this is generalized: the off-board demand at each time step is calculated by starting with the off-board demand
from the prior time step and reducing it by the amount by which the
on-board demand increased. The only other ingredient to understand
the flow of demand in our VRP is that when the truck returns to the
depot node at some time step, it drops off all demand there. The on-board
demand returns to zero, and off-board demands are not changed.

To be explicit, the general update rules of on-board and off-board
demand are

\[
e_{t}=\begin{cases}
0 & \text{if }\xi_{t}=0\\
\min\left(1,e_{t-1}+d_{\xi_{t}}^{t-1}\right) & \text{otherwise}
\end{cases}
\]
and
\[
d_{j}^{t}=\begin{cases}
d_{j}^{t-1} & \text{if }\xi_{t}=0\\
d_{j}^{t-1}-\left(e_{t}-e_{t-1}\right)\delta_{j,\xi_{t}} & \text{otherwise}
\end{cases}
\]
respectively. In these equations, $t$ is a time step index and we
are defining $d_{j}^{t=0}$ to be the demand $d_{j}$. Moreover, $\delta_{kl}$
is the Kronecker delta.

With the definitions above, we can state that the optimization goal
of this basic vehicle routing problem is to find, among all demand-satisfying
routes, the one which completes its route in the smallest time. Here,
time is defined as in equation \eqref{eq:time_basic_vrp}.

\subsection{Tensor Demand Structure\label{subsec:Tensor-Demand-Structure}}

The basic VRP discussed in section \ref{subsec:Basic-Vehicle-Routing}
has the property that all demand must be taken to the same destination
node (the depot). This is built into the mathematical description
of the problem because the off-board demand has a \emph{vector demand
structure}. This means that the demand at a given time step $t$ is
given by a vector 
\[
d^{t}=\left(d_{1}^{t},d_{2}^{t},\ldots,d_{n}^{t}\right).
\]

In a realistic supply chain, we do not have the luxury of a single
delivery destination. Goods from a given node may be split into groups
which need to be taken to various delivery destination nodes. To accomodate
such a situation, we introduce the concept of a \emph{tensor demand
structure}. We begin with a rank-2 tensor.

\subsubsection*{Rank-2 Demand}

Consider a graph with $n$ nodes $z_{1},\ldots,z_{n}$. (There is
no longer any need for a special $z_{0}$ node). We introduce an $n\times n$
matrix $D^{t=0}$ with nonnegative entries. The meaning of the entry
$D_{ij}^{0}$ is, intuitively, the initial amount of demand that is
located at node $i$ and must be shipped to node $j$.

When the truck arrives at node $i$ at time $t$ with this sort of
demand structure, a new issue arises: a \emph{pickup selection }decision
must be made. There are $n$ different sorts of demand that can be
picked up from node $i$: $\left(D_{i1}^{t-1},D_{i2}^{t-1},\ldots,D_{in}^{t-1}\right).$
Because of this difficult issue, we cannot simply determine whether
or not candidate solution giving only a route without an explanation
of how we perform pickup selection is in fact a demand-satisfying
solution.

Assuming that pickup selection can be accomplished in a reasonable
fashion, we can obtain a new matrix $D^{t}$ by reducing $D^{t-1}$
by the amount of demand picked up by the truck from node $i$. However,
now a second issue appears: now that demand is on the truck, we have
to remember which parts of the on-board demand must go to which destination
nodes. This can be dealt with by promoting on-board demand at time
$t$ to a vector $E^{t}=\left(E_{1}^{t},E_{2}^{t},\ldots,E_{n}^{t}\right)$.
The meaning of $E_{i}^{t}$ is that, after the operations at time
step $t$ (including pickup), $E_{i}^{t}$ is the amount of demand
on the truck which must be delivered to node $i$.

Now that there is on-board demand in the truck, we need to revisit
what happens when the truck first arrives at a given node $i$. Before
pickup selection or any other operation, the first step is now to
completely drop off demand $E_{i}^{t-1}$. Mathematically, this simply
means setting $E_{i}^{t}=0$. If we wish, we can also keep track of
the overall total demand satisfied after each time step, in which
case we would iteratively defined a sequence $S$ by 
\begin{align*}
S^{0} & =0\\
S^{t} & =S^{t-1}+E_{\xi_{t}}^{t-1}
\end{align*}
where, as in section \ref{subsec:Basic-Vehicle-Routing}, $\xi_{t}$
refers to the node visited at time step $t$.

\subsubsection*{Arbitrary Rank Demand}

The ideas of a demand matrix $D_{ij}^{t}$ can readily be generalized
to higher-rank tensors. The reason for doing this is that in a practical
supply chain (including the one our study is based on), there are
delivery requirements along the lines of ``move this box from node
3 to node 7, and then from node 7 to node 5''. Such multi-leg requirements
may sound odd, but they can arise for numerous practical reasons.
There may be capacity limitations at node 5, and node 7 may be a storage
warehouse. Or perhaps the box needs to have an operation performed
on it before its final delivery. Another important reason for multi-leg
delivery requirements is that a cargo container may need to be sent
somewhere else after delivery.

Whatever the reason, there promoting the matrix and vector structure
of $D$ and $E$ to higher rank tensors allows us to encode the data
that we need for this new situation. An initial demand tensor $D_{ijk}^{0}$
can be interpreted as ``there is initially demand $D_{ijk}^{0}$
located at node $i$ which needs to first travel to node $j$ and
then travel to node $k$.''

Unfortunately, with higher-rank tensor structure like this, the operations
that are performed when a truck arrives at a node become even more
complicated. Consider first an empty truck arriving at node $i$ at
time $t$. It starts by performing pickup selection to decide what
off-board to pick up: any part of $D_{ijk}^{t-1}$ is fine as long
as the first index is $i$. After the pickup, the off-board demand
is correspondingly reduced. However, the loaded demand is now material
with instructions like ``go to node $j$, then go to node $k$''
so we must introduce a matrix on-board demand $E_{jk}^{t}$ to track
this. However, now that the truck has this on-board demand, when it
later drives to node $j$, the on-board $E_{jk}$ will be dropped
off. This demand is \emph{not satisfied} because it hasn't reached
its final destination of node $k$. We are therefore forced to introduce
a rank-2 matrix off-board demand structure when this demand is dropped
off! When that matrix off-board demand is later picked up, it is converted
to rank-1 vector on-board demand.

In conclusion, rank-$r$ off-board demand will automatically require
tracking off-board demands with ranks 2 through $r$ as well as on-board
demands with ranks $1$ through $r-1$. These can be separately tracked
by a collection of tensors like
\begin{align*}
D^{r\,t} & \quad\\
D^{r-1\,t} & \quad E^{r-1\,t}\\
 & \vdots\\
D^{2\,t} & \quad E^{2\,t}\\
 & \quad E^{1\,t}
\end{align*}
or, alternatively, we can ``embed'' lower rank demand within a single higher rank demand tensor. Regardless of the organizational approach, there
is no question that bookeeping is one of the major issues that arise
when dealing with this more realistic version of a vehicle routing
problem.

As with the cases above, we can introduce a ``total demand satisfied''
sequence $S^{t}$ which accumulates only when demand is sent to its
final destination. We do not accumulate $S$ when rank-2 on-board
deamand arrives at a node, but we do accumulate it when rank-1 on-board
demand arrives because that node is the final destination for that
material.

\subsection{Multiple Trucks}

The next complexity to consider is the involvement of multiple trucks.
This is intuitive and easy to describe mathematically, but it adds
immense difficulty for optimization. The main observation to make
about the mathematical structure is that there is an on-board demand
for every truck, but there is only one off-board demand. Thus, we
need a new index $m$, which ranges from $1$ to the number of trucks
$N$, added to on-board demand. For instance:
\[
E_{m,ij}^{t}
\]
for rank-2 on-board demand. We can allow different trucks to have
different capacities $C_{1},\ldots,C_{N}$, but throughout our work
we assume that all trucks have capacity $1$.

The introduction of multiple trucks adds profound subtleties to the
problem. The optimal solution to an instance with many trucks may
involve highly collaborative relationships between trucks. Finding
nearly optimal solutions may thus require an exploration of numerous
heuristic optimization algorithms or machine learning approaches.

In our study, we take an approach that will not offer a truly collaborative
solution, but may provide good solutions with excellent runtime. We
find truck routes in an iterative fashion, starting with a single
truck, updating the off-board demand expected to be satisfied by that
truck, and then adding a second truck with the new demands. Repeating
this process allows us to take advantage of some of the benefit of
including many trucks, but without dealing with the most intractable
aspects of this problem.

\subsubsection*{Optimization Goals with Multiple Trucks}
\label{sec:multi-truck-optimization-goal}

When dealing with multiple trucks, there is some ambiguity on the optimization goal for a routing problem. There are two reasonable goals to consider.
\begin{enumerate}
\item If the number of trucks is fixed and given, then the optimization goal is to minimize the total driving time among all trucks to satisfy all demand.
\item If the number of trucks is not fixed, then the optimization goal is to minimize total driving time for all trucks and to find the number of trucks that attains the lowest minimal total driving time.
\end{enumerate}
In other words, we can either fix or optimize over the number of trucks.

\subsection{Individual Boxes and the Box Soup Simplification}
\label{sec:box-soup}

Another factor in the realistic vehicle routing problem that we are
building toward is the fact that boxes are not abstract ``material''
but physical boxes with specific volumes and weights and specific
routing requirements. Each box has a starting node and a list of nodes
that it must arrive at before going to its final destination node.
In other words, for every box $a$, where
\[
a\in\{1,\ldots,\text{number of boxes}\},
\]
there is a corresponding list of nodes $R_a$
which box $a$ is required to visit. The number
of elements in $R_a$ is some integer
$r_a$ which is greater than or equal to 2.
\footnote{
We sometimes call $r_a$ the rank of box $a$.}
In other words box $a$ follows the path
\begin{equation}
\label{eq:box-path}
R_{a}=\left(R_{a}^{1},R_{a}^{2},\ldots,R_{a}^{r_{a}}\right).
\end{equation}
There is also a volume for the box $a$ which
we denote as $V_{a}$.

Rather than dealing with all of this detail, we can simplify the problem
by finding all boxes with the same required route $R_{a}$. We can
put all such boxes together into a \emph{box group.} For the optimization
algorithm, there is no need to distinguish boxes within the same box
group unless they have different volumes which will fill off-board
and on-board demand differently. However, we can perform a \emph{box
group soup} simplification where we combine all of the boxes within
the same group together and regard their total volumes as continuous.
The box soup then determines the initial off-board demand tensor(s),
but it plays no other roll after that. In this case, the final solution
we find may not be physically possible, but at least we won't have
to deal with the enormous complexity of tracking individual boxes.

\subsection{Restricted Driving Windows}
\label{sec:driving-window-optimization}
Typical VRPs involve minimizing driving time to accomplish the goal of fulfilling all deliveries. However, in a commercial setting there is a limitation on the time in which trucks can drive. We may fail to satisfy all demand, especially with heuristic algorithms. 

Suppose that all trucks are only allowed to drive during an overall period of time $T_\text{max}$. The trucks drive simultaneously during this time. Then, we are not necessarily guaranteed that it is possible to fully satisfy demand within that constraint. 

In this situation, the objective function is no longer obvious. One possibility is to minimize driving time and maximize satisfied demand with some relative weighting. There is, however, a different commercially natural objective to consider. To minimize, cost, it is important to minimize the number of trucks required. Thus, a useful objective is to find the smallest number of trucks such that it is possible to satisfy all demand.

\subsection{Aisin Corporation Vehicle Routing Problem\label{subsec:full-aisin-vrp}}

With all of the ideas above, we are finally in a position to describe
the commercial routing optimization problem that we aim to address
with an optimization workflow. For brevity, we will refer to the most general (and potentially complicated) of these commercial routing problems as the ``logistical routing problem'' (LRP).

The LRP is based closely on the supply chain of Aisin Corporation, a Japanese automotive manufacturing company. Their operations involve the delivery of parts between numerous facilities. Driving times between these facilities range from minutes to hours, and tens of thousands of boxes must be shipped by trucks between these locations on a daily basis.  The Aisin Corporation routing challenge can be thought of as an instance (or class of instances) of the LRP. For example, the highest rank of demand tensors is 3. When referring specifically to the Aisin Corporation restricted version of the LRP, we will use the terminology ``Aisin logistical routing problem'' or ALRP. The ALRP is the instance for which we describe an implementation of our workflow in section \ref{sec:execution}.

We strongly emphasize that neither the LRP nor the ALRP are meant to be rigorously defined computational problems. There is some ambiguity on the objective goal in these problems. The following is a summary of the features of the LRP and ALRP.

\begin{itemize}
\item \textit{Nodes}: The number of nodes are arbitrary for the LRP. In the ALRP, there are 23 nodes.
\item \textit{Trucks}: Multiple trucks are allowed in the LRP and ALRP. For the ALRP, the typical number of trucks is in the 50-100 range.
\item \textit{Initial Demand}: Off-board demand for the LRP can consist of an arbitrary (finite) number of demand tensors with rank greater than or equal to 2. The ALRP only has rank 2 and rank 3 off-board demand.
\item \textit{Driving Window}: Driving windows are optional for the LRP. In the ALRP, the driving window is fixed (and, in reality, is 16 hours broken into two shifts).
\item \textit{Boxes}: For the LRP, demand can be broken into individual boxes. This is the case for the ALRP.
\item \textit{Time Matrix}: The time matrix is arbitrary for the LRP. The ALRP time matrix is based on actual estimated driving times between facilities in Japan. Times range from minutes to hours in this case.
\item \textit{Objective}: The objective for the LRP/ALRP is ambiguous but is roughly to maximize demand satisfied while minimizing total driving time (which includes minimizing the number of trucks). 
\end{itemize}

\section{Binary Optimization and Annealing\label{sec:Binary-Optimization}}

The annealing algorithms that we discuss below are heuristic algorithms
that attempt to find the minimum or maximum of polynomials with binary
variables. Many problems, especially discrete optimization problems, can be cast as polynomials of binary variables \cite{karp1972reducibility}.  Routing problems like vehicle routing problems are not
manifestly similar to optimizing a polynomial objective function,
but we will be able to find polynomial optimization instances with
solutions that can be used to build up approximate solutions for the
realistic vehicle routing problem discussed in section \ref{subsec:full-aisin-vrp}

\subsection{Polynomials with Binary Variables}

By a ``polynomial with $n$ binary variables'' we mean a polynomial
from\textbf{ }$\{0,1\}^{n}\to\mathbf{R}$ with real coefficients. While
we are choosing the domain of the $n$ variables to be $\{0,1\}^{n}$
there are other reasonable conventions like $\{1,-1\}^{n}$. In the
former case, we sometimes say that the variables are \emph{boolean} and in the
latter case the variables are said to be \emph{spin} variables. When
we use the term binary, without clarification, we mean boolean. As
a example, $f:\{0,1\}^{2}\to\mathbf{R}$ defined by $f(x,y)=3 x y-y$
is a polynomial with two boolean variables.

The computational problem PUBO, which we take to stand for \emph{Polynomial
Unconstrained Binary Optimization}, is the problem of finding the minimizing
input to a given polynomial with binary variables. Similarly, the
computational problem QUBO (\emph{Quadratic Unconstrained Binary
Optimization}) is the same as PUBO except that the polynomial has
at most degree 2. PUBO (and thus also QUBO) are NP-hard, as can be
seen from the fact that many NP-hard problems can be reduced to QUBO \cite{lucas2014ising}.
For the sake of brevity, we often use the term ``QUBO'' to refer
to a quadratic polynomial with binary variables rather than the computational
problem itself as in ``$f(x,y,z)=z+xy-x$ is a QUBO and $g(x,y,z)=xyz$
is a PUBO''.

\subsection{Quantum and Classical Optimization Algorithms}

There has been substantial interest in the QUBO and PUBO problem classes from the quantum
computing community because various heuristic quantum optimization
algorithms are designed polynomials with binary variables. 
The quantum annealers of D-Wave Systems specifically solve QUBO instances. Simulated annealing, which is a class of classical algorithms that are somewhat analogous to the operation of quantum annealing
are well-suited for both QUBO and PUBO instances. 

There are also very notable circuit-model algorithms that optimize
PUBO instances. The quantum approximate optimization
algorithm (QAOA) \cite{farhi2014quantum} is a variational algorithm that
begins with a uniformly distributed quantum state and gradually
evolves the state into one which, when measured, yields an approximate
solution to the original PUBO. In certain respects, QAOA is a natural
choice for our workflow once hardware evolves to the scale where
around 1000 useful qubits, either error-corrected or sufficiently low-noise, can be manipulated in a circuit-model quantum
computer. In lieu of this, the tests performed in section \ref{sec:execution} exclusively use annealing algorithms which we now
briefly review.

\subsubsection{Simulated Annealing}
\label{sec:sim_anneal}

Simulated annealing \cite{pincus1970letter,van1987simulated} refers to a class of heuristic optimization algorithms that are motivated by the
idea of starting in a ``high temperature ensemble'' and gradually
``cooling'' until a state finds its way into the minimum of an objective
function. To be more concrete, consider a finite set of points $A$ and an objective
function $f:A \to \mathbf{R}$. Assume that $f$ is bounded below so that
there is some $a_0 \in a$ such that $f(a_0) \leq f(a)$ for all $a\in A$.
In addition, there is another important piece of structure needed for the
set of points $A$: we need a concept of ``neighboring'' points. Assume,
therefore, that for every $a\in A$, there is a collection of ``neighbors''
of $a$, $N(a)$. While there are no formal requirements on the properties of
$N(a)$, simulated annealing works best when $N(a)$ tends to satisfy some
things that are reasonable for so-called neighbors: we want all of the
neighbors of $a$ to have similar objective values to that of $a$. We
also want the size of $N(a)$ to be quite small compared to the size of
$A$, and we also need to ensure that the elements of $N(a)$ are efficiently
determined given $a$.

PUBO and QUBO fit naturally into this framework. We set $A = \{0, 1\}^n$; neighbors are naturally defined by putting a limitation on Hamming distance.
For example, given a binary string $x = (x_1, \ldots, x_n)$, we can let
the $n$ strings obtained by flipping a single bit of $x$ be the elements of $N(x)$.

The final input needed for simulated annealing is a ``cooling schedule''
which is a value of a ``temperature'' $T$ for every time step. $T$ should
be taken to start off as a larger number and gradually decrease to zero.
We start in an initial guess $a_1 \in A$ and look at all neighbors of $a_1$.
We evaluate the objective function $f(a_1)$ and compare it to $f(a)$ for $a\in N(a_1)$ and we select a new state $a_2$ based on probabilities given by relative Boltzmann factors $e^{-f(a)/T}$ for the various states. When $T$ is larger compared to objective function differences, each step
randomly changes states with no consideration for value of $f$, but as
$T$ tends to zero, states evolve in an increasingly greedy fashion.

\subsubsection{Quantum Annealing}
\label{sec:quantum_anneal}

Quantum annealing is a quantum analog of the classical simulated annealing described above.  And importantly, quantum annealing is the paradigm implemented on the D-Wave Systems quantum annealing computer, which was used in this study to find the solutions to the logistics routing problem instances.  Quantum annealing
is meant as a form of adiabatic quantum computing, where an initial quantum
state is subjected to a time-dependent Hamiltonian in such a way that the
state remains in the (approximate) ground state of the Hamiltonian during
evolution. By evolving from a initial ``driving'' Hamiltonian to a 
``problem Hamiltonian'', one that represents the binary optimization
problem instance that we want to solve, measurements of the final ground
state can yield solutions to the binary optimization problem instance \cite{aharonov2004adiabatic}.

In more detail, a transverse field is applied to an ensemble of spin qubits to provide a high-energy superposition of states in an initial driving Hamiltonian, $H_\mathrm{init}$.  The transverse field serves as an effective high temperature.  The transverse field is slowly lowered, while simultaneously a problem Hamiltonian, $H_\mathrm{final}$ is applied to the qubits via spin-spin couplings and external fields with increasing strength according to an annealing schedule
\begin{equation}
\label{anneal_schedule}
H(s) = (1-s)H_\mathrm{init}(s) + s H_\mathrm{final} ,
\end{equation}
where $s\in[0,1]$ is the annealing schedule parameter.  

If done adiabatically--that is, sufficiently slowly--this allows the final alignment of the qubits, spin up or down, to represent a low-energy minimum configuration that corresponds to a solution to the problem Hamiltonian that was applied.  Even in theory the rate of annealing can be considerably long depending on the minimum gap between energy levels of the Hamiltonian.  Adding to this the limitations of coherence time and noise, adiabatic quantum computing has not yet been achieved.

Quantum annealing is a similar approach to adiabatic quantum computing that uses the similar anneal schedule achieves low-energy solution states without strict adiabaticity \cite{kadowaki1998quantum}.  The intuition behind quantum annealing is that during the anneal quantum effects will allow the exploration of energy structure of the problem by tunneling through energy barriers  \cite{farhi2000quantum,johnson2011quantum,kadowaki1998quantum,santoro2006optimization,denchev2016what}.  In practice, of course, noise limitations limit the effectiveness of the technique, but nonetheless, such devices have been built that increasingly are able to find low-energy states of problem Hamiltonians for ever increasing problem sizes.  

\subsection{D-Wave Systems Quantum Annealing Computer}
\label{sec:dwave}

D-Wave Systems has been offering quantum annealing computers for commercial use for over decade.  Their latest model, the D-Wave Advantage, offers 5000 superconducting qubits operating at cryogenic temperatures.  It is accessible via the Internet and is fully programmable to represent any problem in QUBO formulation of binary variables as described earlier in this section.  In order to overcome the limitation of problem size--a limitation for all existing quantum computing devices of today--D-Wave Systems include a hybrid solver that uses the quantum annealing computer to provide promising starting points to large-scale classical computers.  This allows researchers to work on the quantum annealing algorithm while running much larger problem sizes of practical interest.  This hybrid solver was in fact used in this study to address a much larger and more realistic logistics problem than could be handled by just the quantum annealer alone.

\subsubsection{D-Wave Quantum Annealer}
\label{sec:dw-quantum_anneal}

The important characteristics of a quantum annealer are the number of qubits, the amount of connectivity among the qubits, and the noise qualities of the qubits.  The latest D-Wave Advantage also offers an increase in couplers between qubits of 16, up from 6 on previous versions.  This increase in connectivity allows more efficient embedding of logical problems onto the hardware graph of qubits, and thus enables larger problem sizes to be embedded onto the processor.  The embedding of the logical problem onto the hardware qubit graph is important.  One should recognize that the 5000-qubit processor cannot handle a problem of 5000 binary variables.  The embedding requires multiple hardware qubits to be programmed as a logical node to represent each logical variable.  For a fully-connected logical problem, one in which every binary variable interacts with all the others, one can only embed such a fully-connected problem of approximately 180 logical binary variables.  Many problems of practical interest are not fully-connected logically, so larger problem sizes of hundreds of binary variables can often be embedded.  However, to handle much larger problems of thousands of binary variables requires the use of the hybrid solver--quantum annealer and classical heuristic solvers working together. A summary white paper fully describing the characteristics and operating modes of the D-Wave Advantage quantum annealing computer can be found on their web site \cite{dwave2017advantage}. 

The performance of the D-Wave Systems quantum annealers has increase with each new generation of machine.  While the latest machine still can not perform better than the best classical algorithms on multiple CPU/GPU compute hardware, they are narrowing the gap between their annealing QPU and a classical CPU for certain problem types  \cite{denchev2016what}.  It is difficult to compare the performance of quantum annealing and classical heuristics in theory.  Predictions of adiabatic quantum computing have some theoretical underpinning, quantum annealing itself does not.  Additionally, in empirical comparisons, the quantum annealers are mostly limited by noise and other imperfections in their current state of technology, while classical algorithms and hardware are very mature.  Research on improving quantum annealing is ongoing both from a qubit technology perspective and in an operational perspective.  A case in point for the latter is exploring tailored annealing schedules different than the nominal linear annealing schedule shown above in Equation \ref{anneal_schedule}.  For example, inhomogeneous annealing schedules for each qubit can be applied and have shown significant improvements in obtaining higher probabilities for low-energy solutions. \cite{adame2018inhomogeneous}

\subsubsection{D-Wave Hybrid Solver}
\label{sec:dwave-hybrid}

The D-Wave Systems hybrid solver aims to bring quantum computing to bear on larger problem sizes than it can alone handle at this point in its maturity. The key idea is to use the quantum annealer to better guide a classical heuristic algorithm.  Heuristic algorithms for optimization lie between the two extremes of exhaustive search and random sampling.  Exhaustive search is, of course, too time consuming for large problems, scaling exponentially with problem size for combinatorial optimization.  Random sampling is fast but not very good at finding extrema.  Heuristics come extend from each of these extremes.  Examples are Search techniques that can smartly eliminate large portions of the search space, such as in branch and bound, and random sampling along with local search, such as in simulated annealing described above.  For discrete optimization, which cannot benefit from the power of gradient descent, alternative heuristics have been developed that can help guide the search space or eliminate redundancy, such as tabu search that excises previously explored territories of the search strategy.

Quantum annealing brings something additional into play.  It is believed, if not yet proven, that quantum annealing will be better at exploring the global search space and, via the power of quantum tunneling, might avoid getting stuck in local minima.  The D-Wave Systems hybrid approach invokes several of these approaches in an overarching meta-heuristic \cite{dwave2021hybrid}.  

\begin{figure}
\centering
\includegraphics[width=.5\textwidth]{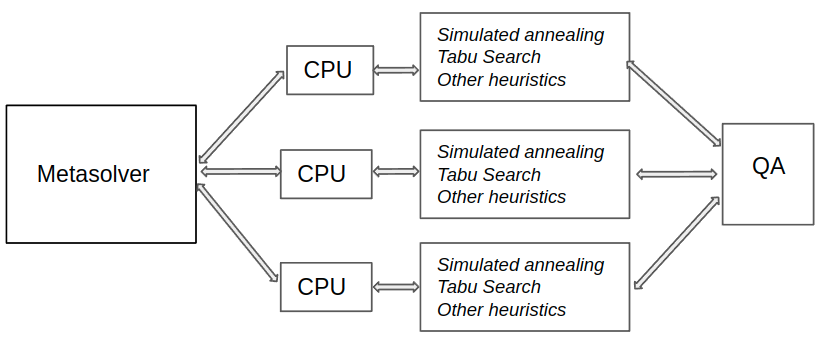}
\caption{A functional diagram of the D-Wave Systems hybrid solver.  The Metasolver on the left governs the overarching algorithm.  It launches multiple threads which uses classical heuristic algorithms running on CPUs or GPUs as depicted in the middle of the figure.  Simultaneously, it uses the quantum annealer on the right to search for promising solutions of smaller subsets of the problem to provide additional promising starting points to the heuristic algorithms.}
\label{fig:dwave-hybrid}
\end{figure}

A top-level functional diagram of the hybrid solver is shown in Figure \ref{fig:dwave-hybrid}.  The Metasolver on the left governs the overarching algorithm.  The problem Hamiltonian in QUBO form is provided along with the stopping conditions, usually just an overall target run-time.  The Metasolver launches multiple threads which uses classical heuristic algorithms running on CPUs or GPUs.  These heuristic algorithms use versions of simulated annealing, taboo local searches, and additional proprietary heuristics.  Simultaneously, the Metasolver uses the quantum annealer to search for promising solutions of smaller subsets of the problem, and these are fed back into the algorithm flow to provide additional promising starting points to the heuristic algorithms in each thread.  The Metasolver collects a set of best solutions until the stopping condition is reached and the results are reported back to the user.

In performance testing on benchmark optimization problems, the hybrid solver performs well and in some cases better than other state-of-the-art classical heuristics on comparable hardware.  Additionally, when running the hybrid solver with the quantum annealing augmentation disabled, the solution time and quality are degraded slightly on average.  This indicates that the quantum annealing solutions are providing improved performance, at least for the problem types and instances studied to date.  The long-term expectation is that as the quantum annealing hardware improves in the future, this should provide even more advantage over the purely classical solvers alone.  Interestingly, this approach can also be pursued using other quantum computing paradigms, such as circuit model discrete optimization solvers.  Further testing will need to be done on future quantum hardware and quantum algorithms--both quantum annealing and circuit model--to see how much further these hybrid approaches can be developed.

\section{Single Truck PUBO\label{sec:Single-Truck-PUBO}}

The enormously complex logistical routing problem (LRP) described in section
\ref{subsec:full-aisin-vrp} could be formulated as the problem of
minimizing a polynomial with binary variables. However, such a formulation
would involve an intractably large number of variables and may have
very high polynomial degree. This complexity suggests that we find another approach, even if that approach cannot achieve an exactly optimal solution.

Consider a single truck and ignore completely the presence of all
other trucks. We do not concern ourselves with the fact that other
trucks can potentially interfere with or assist this one truck. We
only focus on one single truck and we write down a PUBO, the solution
to which ought to give a reasonable route for this truck.

\subsubsection*{Binary Variables}

Suppose that there are $n$ nodes $z_{0},\ldots,z_{n-1}$. (Note that
we are starting indices at 0 for convenience here. $z_{0}$ is not
a special node as in section \ref{subsec:Basic-Vehicle-Routing}.)
The truck starts at one of the nodes at the first time step $t=0$
and then drives to another one at each subsequent time step. Assume
that there are a total of $\tau$ time steps so that the route of
the truck is $\xi_{0},\xi_{1},\ldots,\xi_{\tau-1}$ with each $\xi_{t}\in\{0,1,\ldots,n-1\}$.
The fact that we are fixing the total number of time steps will play
an important role below.

We now introduce $n\,\tau$ boolean variables
\[
\left\{ x_{it}\,|\,i\in\{0,1,\ldots,n-1\},t\in\{0,1,\ldots,\tau-1\}\right\} .
\]
The intended meaning of these variables is that $x_{it}$ is supposed
to be $1$ if the truck is located at node $i$ at time
step $t$ and $x_{it} = 0$ otherwise. This approach to introducing binary variables is commonplace
when dealing with PUBO formulations of routing problems like the traveling
salesman problem.

\subsubsection*{Locality Term}

Configurations of the binary variables $x_{it}$ can violate locality:
if $x_{2\,5}=x_{3\,5}=1$, then the truck is apparently at $z_{2}$
and $z_{3}$ simultaneously at time step $5$. For this reason, the
PUBO will require terms that act as constraints to enforce locality.
We must demand that 
\[
\sum_{i}x_{it}=1
\]
for all times $t$. There are a few ways to handle this, but our approach
is to start our PUBO with a quadratic term of the form
\begin{equation}
f_{\text{local}}(x)=A_{\text{local}}\sum_{t}\left(1-\sum_{i}x_{it}\right)^{2}\label{eq:local-term}
\end{equation}
which is 0 if and only if the configuration is local at all times.
The coefficient $A_{\text{local}}$ is some positive real number which
can be tuned to improve an algorithm's performance. $f_{\text{local}}(x)$
is positive when the configuration is nonlocal and it gets larger
as the solution becomes more and more non-local. $f_{\text{local}}$
also prohibits configurations where a all variables are zero at any
given time.

\subsubsection*{Demand and Time}

The tensor demand structure of the LRP explained in sectio \ref{subsec:Tensor-Demand-Structure}
posses a challenge for a PUBO formulation. Off-board demand changes
as trucks pick up and drop off boxes, and we thus need a dynamical
approach to deal with demand to be technically correct. However, the
single truck PUBO we are building is a heuristic, so we allow a heuristic
approach at this point.

Consider rank-2 and rank-3 off-board demand$D_{ij}^{r=2}$ and $D_{ijk}^{r=3}$.
(To review, the rank-3 demand means material which starts a node $i$
and must be brought to node $j$ and then to node $k$.) We can use
this demand to compute an ``overall off-board demand'':
\begin{equation}
\overline{D}_{ij}=D_{ij}^{r=2}+\sum_{k}\left(D_{ijk}^{r=3}+D_{kij}^{r=3}\right)\label{eq:overall-demand}
\end{equation}
this overall demand can be interpreted as an upper bound on the amount
of material that a truck could encounter at node $i$ that needs to
be sent to node $j$. We then introduce a demand term for the PUBO:
\begin{equation}
f_{\text{demand}}(x)=-A_{\text{demand}}\sum_{ijt}\sum_{\delta=1}^{\delta_{*}(t)}\overline{D}_{ij}x_{i\,t}\,x_{j\,t+\delta}\label{eq:demand-term}
\end{equation}
where $\delta_{*}(t)=\min(\delta_{\text{max}},\tau-1-t)$ and $\delta_{\text{max}}$
is some fixed integer limitation on $\delta$. The reason that we
include $\delta>1$ terms is that a truck may pick something up from
node $i$ with destination at node $j$, but the truck might stop
at node $k$ before going to node $j$. In this case, we still want
to reward the truck from driving from $i$ to $j$ eventually. Solutions
which approximately minimize $f_{\text{demand}}$ may correspond to
routes where the truck can carry large amounts of demand.

We can similarly penalize routes that take excessively long drives.
If $T_{ij}$ is the time matrix computing the driving time between
nodes, adding 
\begin{equation}
f_{\text{time}}(x)=A_{\text{time}}\sum_{ijt}T_{ij}x_{i\,t}x_{j\,t+1}\label{eq:time-term}
\end{equation}
to the PUBO will add a preference for routes that can visit more nodes
in the limited number of time steps. Note that terms \eqref{eq:demand-term}
and \eqref{eq:time-term} can be combined into one quadratic term with
modified coefficients.

\subsubsection*{Redundancy Abatement}

The final term to build our PUBO requires some explanation. Consider
again the ``overall demand'' matrix $\overline{D}_{ij}$. There
are two situations to consider: when $\overline{D}_{ij}$ is large
and when it is small. When it is large, our usage of $\overline{D}_{ij}$
is quite reasonable. Configurations that try to minimize $f_{\text{demand}}$
and $f_{\text{time}}$ are likely to satisfy lots of demand while
avoiding long times. However, in the case where $\overline{D}_{ij}$
is small we encounter a serious deficiency. As mention above, off-board
demand, when treated without any simplification, is dynamical: it changes
as trucks pick up and drop off parts.

Consider the following extreme example. Suppose that between nodes
$z_{3}$ and $z_{6}$, there is a small amount of demand $\overline{D}_{3\,6}$.
If our truck drives from $z_{3}$ to $z_{6}$ once, doing so a second
time would be pointless. However, there is no information that this
would be wasteful in $f_{\text{demand}}$. In fact, $f_{\text{demand}}$
may suggest that going back and forth between $z_{3}$ and $z_{6}$
for all time steps is a good solution! To avoid this sort of thing,
we introduce terms in the PUBO which penalize route repetition. However,
we only want to do this for pairs of nodes where $\overline{D}_{ij}$
is ``small'' so we will only introduce these new terms for certain
pairs of nodes. We will use the notation $I_{ij}=1$ when $\overline{D}_{ij}$
is sufficiently ``small'' that we would want to include such a term
and we will write $I_{ij}=0$ for ``large'' $\overline{D}_{ij}$.

In the example above, including a term proportional to 
\[
\sum_{t=0}^{\tau-4}x_{3\,t}\,x_{6\,t+1}\,x_{3\,t+2}\,x_{6\,t+3}
\]
counts the number of times that a truck follows the path $z_{3}\to z_{6}\to z_{3}\to z_{6}$.
We can generalize this term further by adding a delay $\delta\in\{2,\ldots\}$
between repetitions as follows:
\[
h_{\delta}(3,6)=\sum_{t=0}^{\tau-2-\delta}x_{3\,t}\,x_{6\,t+1}\,x_{3\,t+\delta}\,x_{6\,t+1+\delta}
\]
which is the same as before except now we are counting the number
of times that the truck follows the two-step path $z_{3}\to z_{6}$
at some time $t$ and then later follows the same path at time $t+\delta$.
We can then define
\[
h(i,j)=\sum_{\delta=2}^{\tau-2}\sum_{t=0}^{\tau-2-\delta}x_{i\,t}\,x_{j\,t+1}\,x_{i\,t+\delta}\,x_{j\,t+1+\delta}
\]
Finally, we can give the redundancy abatement term for our PUBO
\begin{equation}
f_{\text{nonredundant}}^{I}(x)=A_{\text{nonredundant}}\sum_{ij}I_{ij}\ h(i,j)\label{eq:no-repeat}
\end{equation}
where $I$ is an $n\times n$ matrix consisting of only zeros and
ones which is supposed to be 1 for pairs $ij$ if and only demand
$\overline{D}_{ij}$ is ``small'' enough to include such a repetition-avoidant
term. Note that equation \eqref{eq:no-repeat} is fourth-order, and
we should thus use these terms sparingly: $I_{ij}$ should be zero
unless we expect cycling between $i$ and $j$ to be a serious problem.

\subsubsection*{Full Single Truck PUBO}
\label{sec:full-single-truck-pubo}

Combining the ideas from above, our full single-truck PUBO is the
summation of equations \eqref{eq:local-term}, \eqref{eq:demand-term},
\eqref{eq:time-term}, and \eqref{eq:no-repeat}. There are four positive
constants that can be tuned: $A_{\text{local}},A_{\text{demand}},A_{\text{time}},$
and $A_{\text{nonredundant}}$ as well as the choice of $I_{ij}$ and $\delta_{\text{max}}$.
These parameters offer substantial freedom to tune heuristic PUBO-solvers
to obtain better solutions for the final combined problem instance.

\subsubsection*{Solution Rectification}
\label{sec:rectification}
Approximate solutions to a single-truck PUBO do not always give a well-defined route for a truck. It's possible that a solver returns a solution which is not a minimum of $f_\text{local}$. For these situations, it's helpful to settle on a method for ``rectifying'' solutions that violate locality. 

To enforce locality on an approximate solution $x$, we iterate through time steps $t$. If we have $x_{it} = 0$ for all $i$, then we randomly choose one node $i_*$ to have $x_{i_* t}=1$ while keeping  $x_{it} = 0$ when $i\neq i_*$. 

If, for some $t$, our initial solution has multiple nodes $i_1,i_2,\ldots, i_k$ with $x_{i_s t}=1$, then we wish to select a single value of $s_* \in \{1,\ldots, k\}$ to keep $x_{i_{s_*} t}=1$ while setting the other variables equal to 0. To do this, we simply look at the terms of the PUBO that involve this time step $t$ and we try the $k$ possibilities by brute force. We select any of the choices with the lowest objective function value.

\subsubsection*{Order Reduction}
\label{sec:order-reduction}
To submit a PUBO on D-Wave Systems solvers, including
their quantum annealer as well as simulated annealing and
the D-Wave hybrid solver, there is a necessity to eliminate
any polynomial terms with order exceeding two.

Polynomial order can be reduced at the cost of introducing
extra variables and constraints. This is easy to see through an example. Given a sufficiently large real number $\lambda$,
the polynomial of binary variables $f(x,y,z) = x y z$ is ``equivalent'' to $g(x, y, z, a) = a z + \lambda (a - x y )^2$ in the sense that the minimum of $g$, restricted to the variables $(x, y, z)$, is also a minimum of $f$. To see this, note that the variable $a$, being binary, takes on the same values as $xy$ (0 and 1 only). Moreover, the term $\lambda (a - x y )^2$ enforces
the constraint that $a = xy$ to minimize $g$ as long as $\lambda$ 
is large enough.

In the executions discussed in section \ref{sec:execution},
we perform PUBO  to QUBO conversions for simulated annealing
and for D-Wave hybrid runs. For all cases, we used the
open-source library Qubovert \cite{qubovert} to perform PUBO to QUBO conversions.

\section{Supply Chain Workflow
\label{sec:Supply-Chain-Workflow}}

An approximate solution to the single-truck PUBO described of section
\ref{sec:Single-Truck-PUBO} gives a route for one truck, but the
logistical routing problem (LRP) of section \ref{subsec:full-aisin-vrp}
involves many trucks. Constructing a PUBO with solutions that obtain
multi-truck solutions for a multi-truck VRP is possible, but will
involve a large number of variables and terms. We certainly do not
consider such an approach to be viable with hardware such as quantum
annealers constructed by D-Wave Systems.

The single-truck PUBO, on the other hand, can viably be attacked with near-term quantum algorithms. We thus devise an algorithm which attempts to solve the full LRP by working truck-by-truck, iteratively obtaining a route, estimating the demand satisfied by the route, and then updating the remaining off-board demand tensor before proceeding to the next truck. 

For this truck-loop algorithm, we work only with ``overall demand'' given in equation \eqref{eq:overall-demand}. This demand is a useful simplification to avoid the challenges of higher-rank demand while still obtaining a reasonable solution. We return to the details of higher-rank demand in section \ref{sec:full-scale-sim}.

The truck-loop algorithm is given in pseudocode as follows:

\begin{algorithm}[H]
\caption{Truck Loop}

\begin{algorithmic}
\STATE Inputs:
\STATE \quad Number of time steps to use $\tau$
\STATE \quad Stopping condition \code{stop}
\STATE \quad Solver with rectification \code{solver}
\STATE \quad Initial overall demand $D$
\STATE \code{routes} $\gets$ empty list
\STATE $\ell \gets 0$

\WHILE {not $\code{stop}(D, m)$}
  \STATE $f \gets \code{single\_truck\_pubo}(D, \tau)$
  \STATE $\xi \gets \code{solver}(f)$
  \STATE $D \gets D - \code{estimate\_demand}(D, \xi)$
  \STATE $\code{routes} \gets \code{append}(\code{routes}, \xi)$
\STATE $\ell \gets$ length of \code{routes} 
\ENDWHILE
\RETURN \code{routes}

\end{algorithmic}
\end{algorithm}

This algorithm refers to some subroutines that require explanation:
\begin{itemize}
\item \code{single\_truck\_pubo} returns a PUBO as described in section \ref{sec:full-single-truck-pubo}. The only part of this function that cannot be read-off from section \ref{sec:Single-Truck-PUBO} is the choice of redundancy abatement coefficients $I_{ij}$. These are chosen to be $0$ for a pair $i,j$ when $D_{ij}$ is above some threshold and $1$ otherwise.

\item \code{solver} refers to some PUBO solver and post-processor. The PUBO solver can be any exact or heuristic algorithm which is aims to find the minimum of a given PUBO. The output of \code{solver} is not, however, the $\min$ or $\text{argmin}$ of the PUBO. Instead, we first get the binary variable assignments that approximately minimize the PUBO, and then we perform locality rectification as explained in section \ref{sec:rectification}. 

Moreover, if there are driving windows (see section \ref{sec:driving-window-optimization}), then we also cut off the last steps of the route until the route fits into the window. Along the same lines, we may also lengthen routes that fall well-short of the driving window. This can be done by looping the original route until adding one more step would exceed the time window. This procedure may sound counterproductive, but in applications with driving windows, as in the ALRP (section \ref{subsec:full-aisin-vrp}), we may calculate cost by assuming that all trucks have a fixed price for a fixed window, and in that case it's ideal to use the truck throughout the time window.

Finally, the output of \code{solver} is the sequence of nodes for the single-truck route obtained.  

\item \code{estimate\_demand} is a function which attempts to estimate the degree to which $D$ will be reduced by a truck following a given route. We compute this by performing a small supply chain simulation for the route. Suppose that the truck starts at node $z_4$ and that it will then go to nodes $z_6, z_8, z_3$ in that order. We then load onto an abstract truck the largest allowed amount of $D_{4,6}$ (limited either by the truck capacity or by the value of $D_{4,6}$. If there is remaining truck capacity we move on to $D_{4,8}$, and so on until we either take all of $D_{4,3}$ or we run out of capacity. At this point, the truck (abstractly) drives to $z_6$ and we unload all of the $D_{4, 6}$ that was on-board. We then repeat the procedure starting with $D_{6, 8}$ followed by $D_{6, 3}$. This method provides a reasonable estimate of the demand that a truck is expected to satisfy, but it is only a heuristic approach, severely weakened by the realities of multi-truck interactions.

\item \code{stop} is a stopping condition such that $\code{stop}(D, m)$ is true if and only if conditions on a demand matrix $D$ and the number of trucks assigned $m$ are satisfied. We can, for example, stop whenever more than 50 trucks have been assigned or when all entries of $D$ are below some threshold.
\end{itemize}

Once the truck loop algorithm is complete, we obtain a collection of routes for each truck 
\begin{equation}
\xi = \left\{\xi_m\, | \, m=1,\ldots, N\right\}
\label{eq:truck-loop-output}
\end{equation}
 where $N$ is the number of trucks and each $\xi_m$ is a list of nodes $\left(\xi_{m,1}, \xi_{m,2}, \ldots, \xi_{m,k_m}\right)$ with $k_m \leq \tau$ for all $m$. Unfortunately, this final routing was obtained with a series of heuristics and it's therefore not immediately obvious how to estimate the solution quality. The most naive and easily implemented strategy is to simply look at the final value of $D$ in the algorithm and the total time for all of the trucks to drive along the routes. However, this demand and time calculation is only a approximation. To determine the quality of the solution, a more involved supply chain simulation is necessary, and we explain this in section \ref{sec:full-scale-sim}.
 
 \section{Execution and Performance}
 \label{sec:execution}

 \begin{figure}
\centering
\includegraphics[width=.5\textwidth]{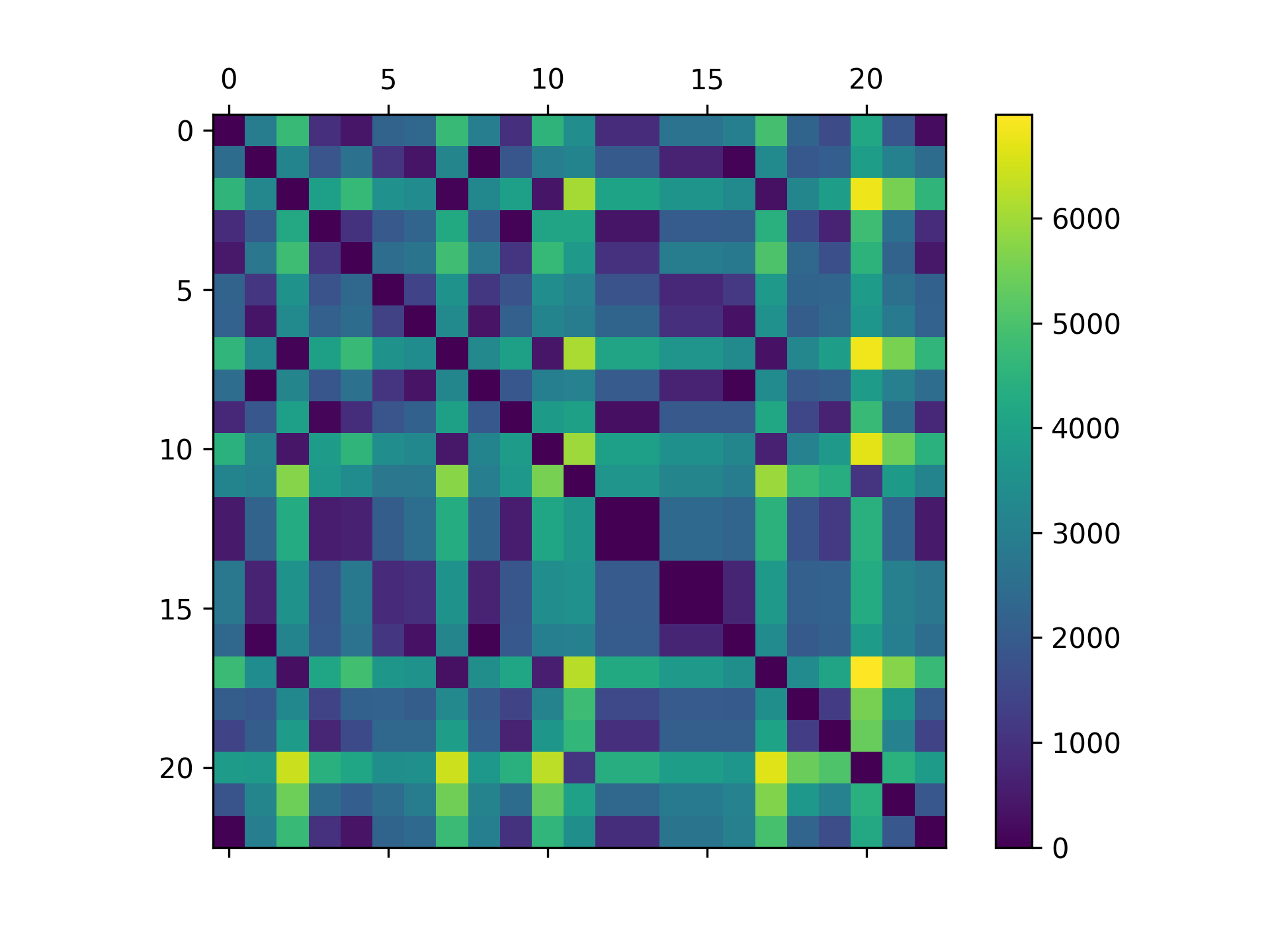}
\caption{The matrix of driving times between the 23 nodes in the Aisin logistical routing problem. Units are in seconds which is the same unit used for the PUBO construction.}
\label{fig:time-matrix}
\end{figure}

\begin{figure}
\centering
\includegraphics[width=.5\textwidth]{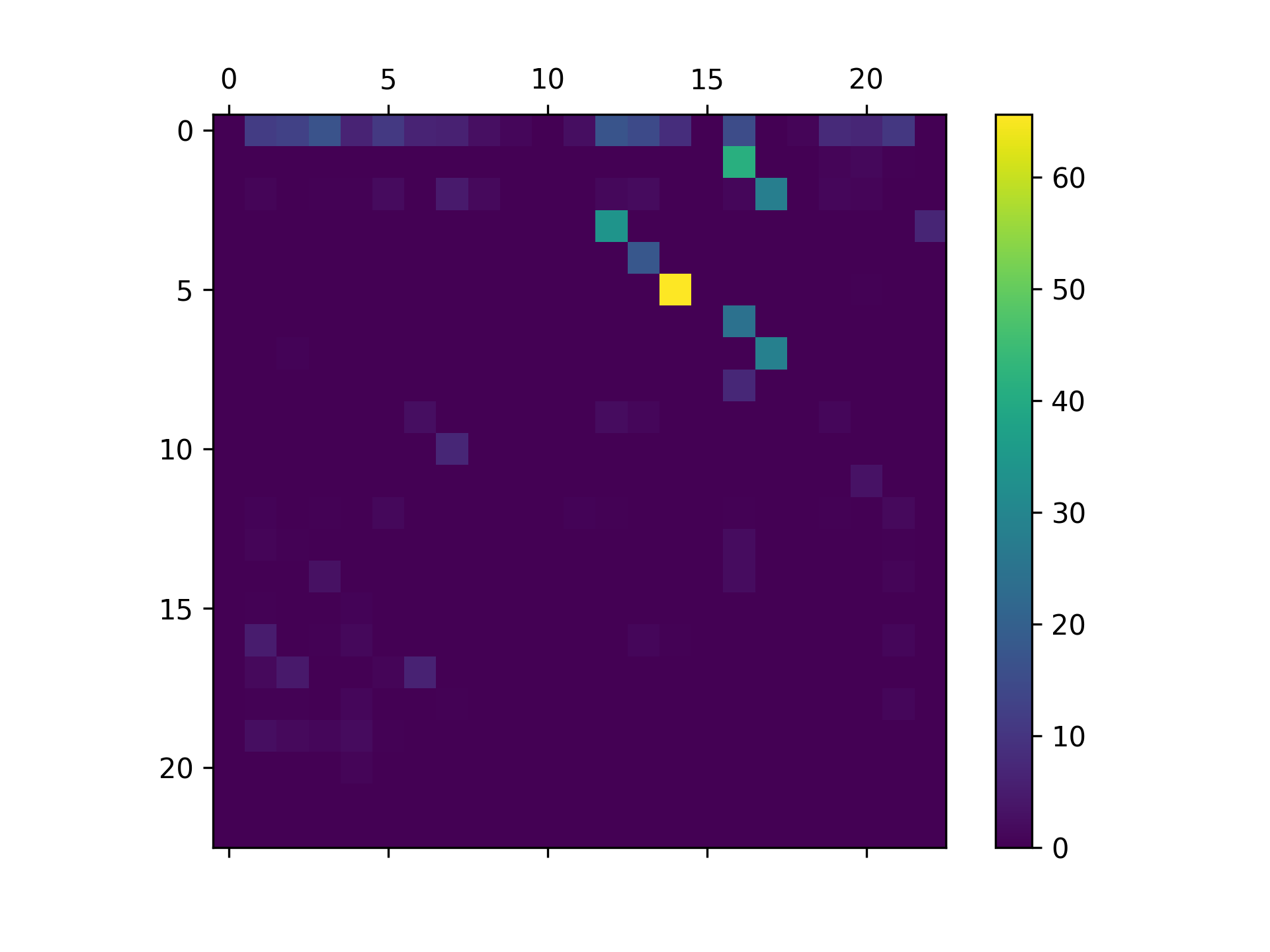}
\caption{The overall demand matrix (as defined in equation \eqref{eq:overall-demand})
for the Aisin Logistics Problem of section \ref{subsec:full-aisin-vrp}. Note that of the $23^2 = 529$ entries for the matrix, only 115 are nonzero.}
\label{fig:demand-matrix}
\end{figure}

 The purpose of the truck loop algorithm of section \ref{sec:Supply-Chain-Workflow} is to have an algorithm that can break a very large commercial problem into sufficiently small pieces that simulated and quantum annealing algorithms can viably be applied in the near term. As a proof-of-concept, we applied our algorithm to proprietary supply chain data of Aisin Corporation using both simulated annealing and the D-Wave Hybrid algorithm. Application of direct quantum annealing rather than the D-Wave Hybrid algorithm remains unrealistic for this problem until the hardware matures to some extent.
 
 We referred to the Aisin problem was referred to as the ALRP in section \ref{subsec:full-aisin-vrp}. Their data consists of approximately 350,000 boxes to be shipped by trucks among 23 nodes in Japan. The travel times between nodes is shown in figure \ref{fig:time-matrix}.The boxes have rank-2 and rank-3 demand structure. Thus, some boxes have direct delivery requirements and others have a required stop along the way. To apply the truck loop algorithm, we first group boxes by their required routes and then we apply the ``box soup'' simplification of section \ref{sec:box-soup} to convert from a discrete to continuous demand structure. Moreover, we use equation \eqref{eq:overall-demand} to compute the overall demand matrix. With truck volumes normalized to 1, the overall demand matrix is illustrated in figure \ref{fig:demand-matrix}.
 
\subsection{Truck Loop Annealing Runs}
\label{sec:truck-loop-execution}

\begin{figure}
\centering
\includegraphics[width=.45\textwidth]{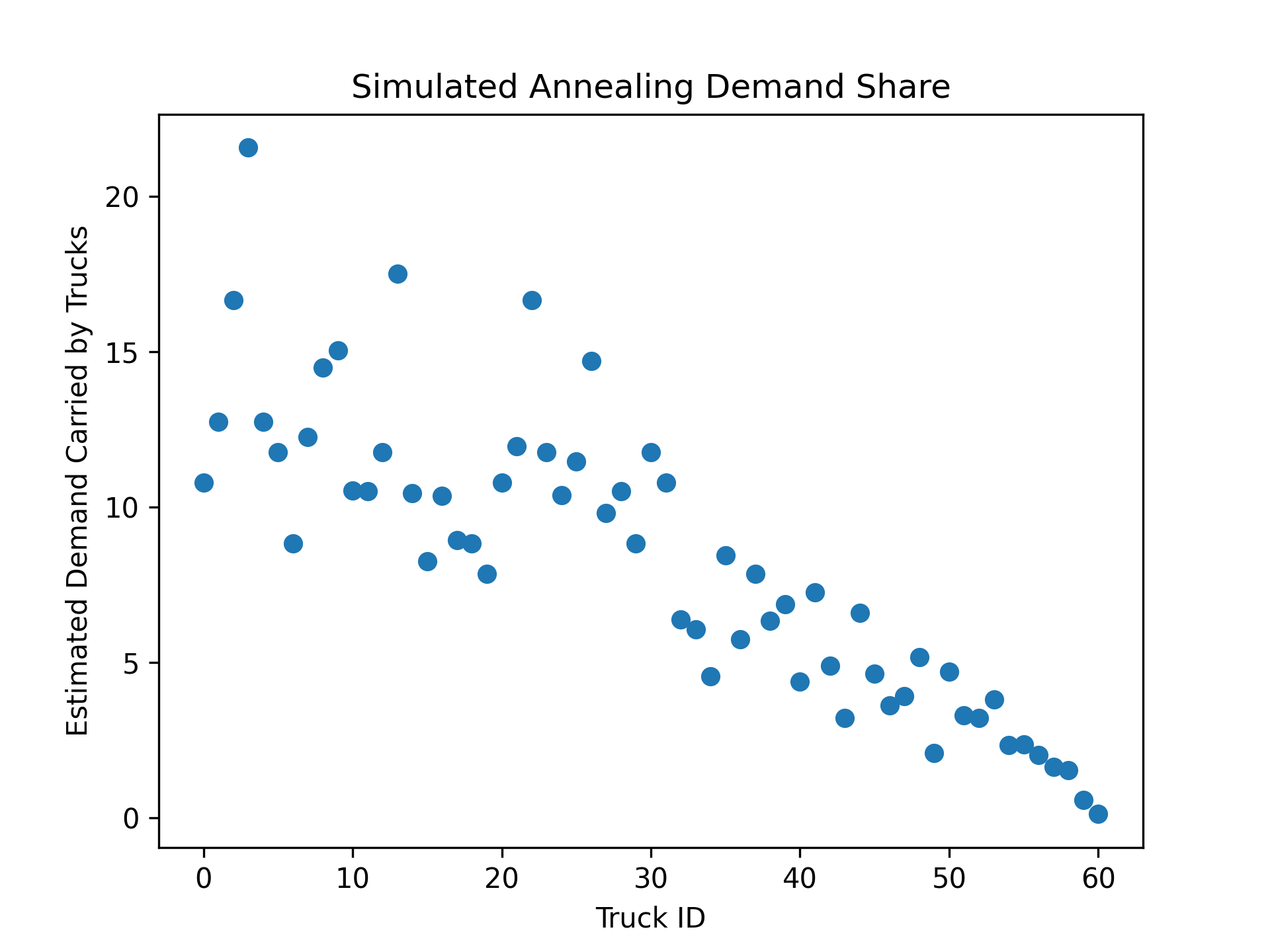}
\caption{Estimated demands computed during the truck loop process for simulated annealing. The algorithm terminated after 61 trucks were given routes. Demands are estimated by the method explained in section \ref{sec:Supply-Chain-Workflow}.}
\label{fig:simulated-annealing}
\end{figure}

\begin{figure}
\centering
\includegraphics[width=.45\textwidth]{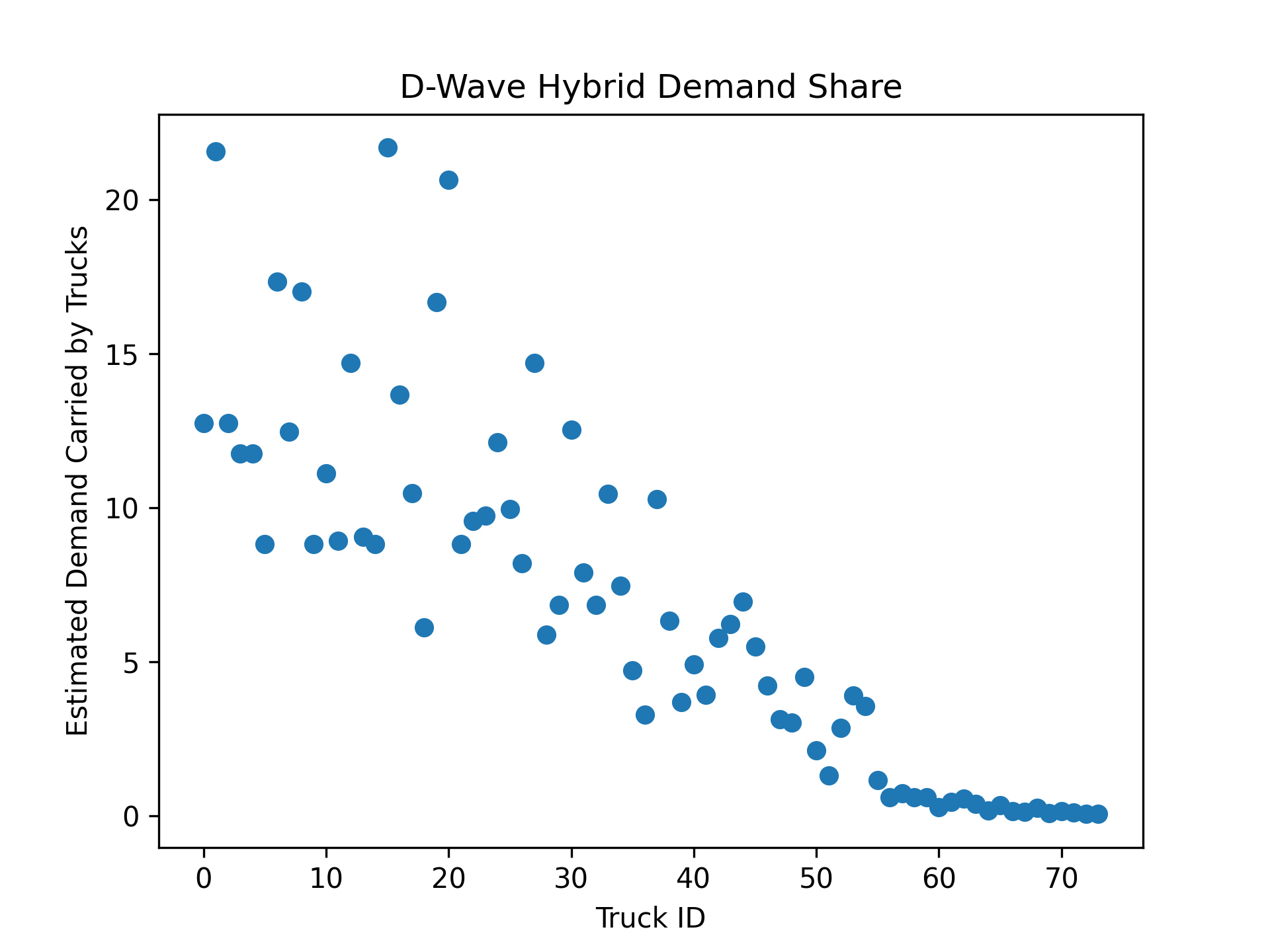}
\caption{Estimated demands computed during the truck loop process using the D-Wave Hybrid solver. The algorithm terminated after 74 trucks were given routes. Note that the last 15 trucks consume a very small amount of demand, suggesting that they can be replaced by a smaller number of trucks. Demands are estimated by the method explained in section \ref{sec:Supply-Chain-Workflow}.}
\label{fig:hybrid}
\end{figure}

We ran the truck loop algorithm of section \ref{sec:Supply-Chain-Workflow} using both a simulated annealing algorithm (section \ref{sec:sim_anneal}) and the D-Wave Hybrid solver (section \ref{sec:quantum_anneal}).

During each step of the loop, we computed an estimated overall demand carried by each
truck. That estimated overall demand is shown for the case of simulated annealing and D-Wave Hybrid in figures \ref{fig:simulated-annealing} and \ref{fig:hybrid} respectively. The simulated annealing loop terminated after 61 trucks, and the D-Wave Hybrid loop terminated after 74 trucks. This termination condition, defining \code{stop} from section \ref{sec:Supply-Chain-Workflow}, is simply to stop all entries of $D$ go under a very small cutoff ($.0005$ in our case).

To construct the single-truck PUBO, we used a different value of $\tau$ (the number of time steps) for simulated annealing and for the D-Wave Hybrid solver: 15 for simulated annealing and 5 for D-Wave Hybrid. With 15 time steps, the typical number of PUBO variables is approximately 350, corresponding to approximately 2500 QUBO variables. With only 5 time steps, there are around 100 PUBO variables which are equivalent to about 200 QUBO variables.

The PUBO coefficients were the same in both cases:

\begin{align*}
    &A_{\text{local}} =  5000,\\
    &A_{\text{demand}} = 320,\\
    &A_{\text{time}} = .01,\\
    &A_{\text{nonredundant}} = 1.
\end{align*}

As we emphasized earlier, the data in figures \ref{fig:simulated-annealing} and \ref{fig:hybrid} cannot, on their own, be taken as an evaluation of the performance of our methodology.  This is especially true given that demands are only estimated by the methods explained in section \ref{sec:Supply-Chain-Workflow}. We now explain how we translated the routes and heuristic solutions from this section into meaningful instructions for a commercial supply chain and, in doing so, evaluate the performance of our workflow.

 \subsection{Full-Scale Simulation}
 \label{sec:full-scale-sim}

 \begin{figure}
\centering
\includegraphics[width=.45\textwidth]{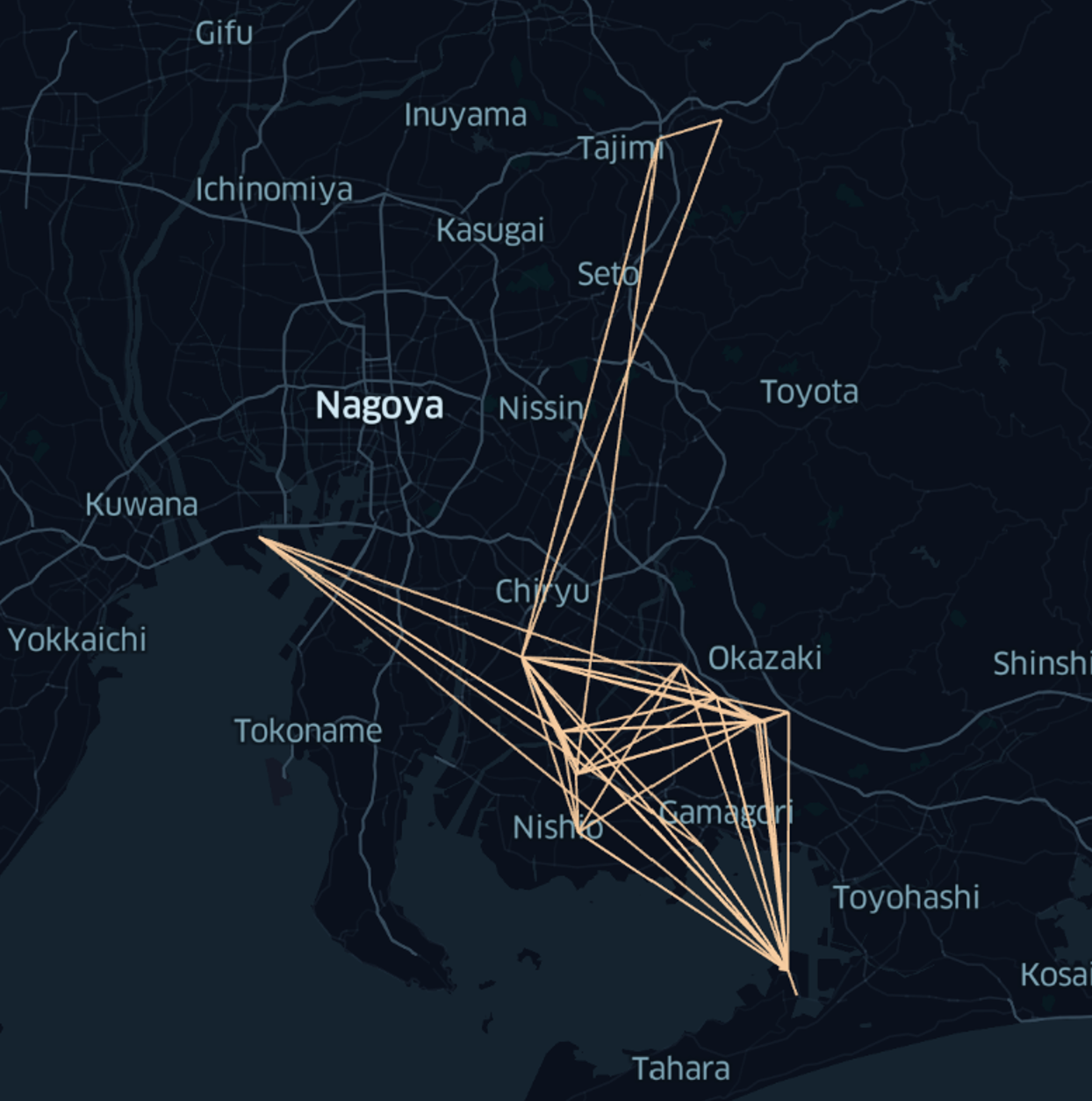}
\caption{Connectivity graph for routes currently used by Aisin Corporation. Lines indicate pairs of the 23 nodes that trucks drive between in the routes determined
by Aisin Corporation logistics experts. For this routing, 142 trucks deliver
approximately 340,000 boxes containing $\sim 15,000$ unique parts. Among these parts, there are 115 unique routing requirements as described in section \ref{sec:box-soup}.}
\label{fig:route-map}
\end{figure}
 \begin{figure}
\centering
\includegraphics[width=.45\textwidth]{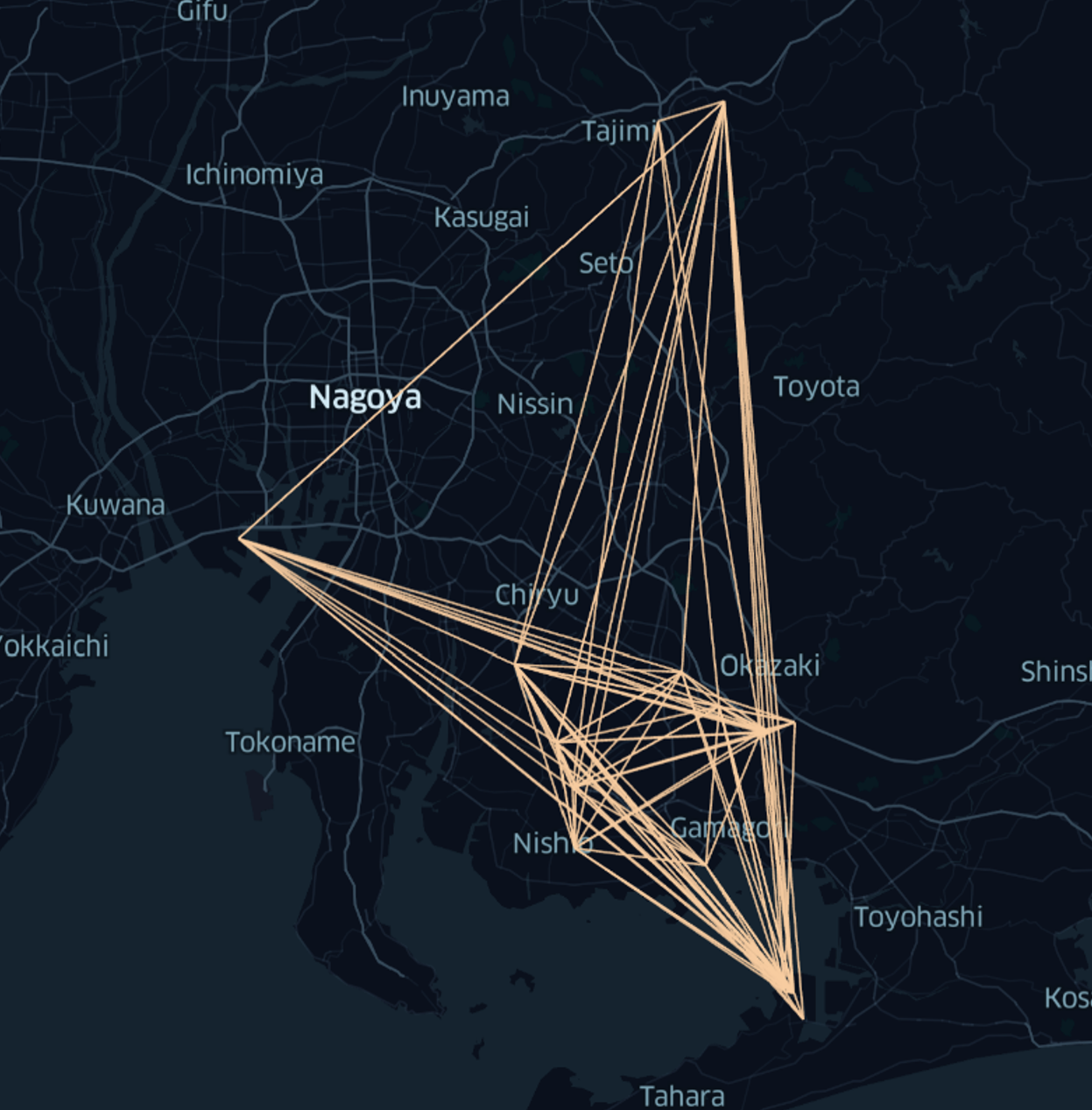}
\caption{Connectivity graph for routes found through our workflow with 61 trucks, the D-Wave hybrid solver, and full-scale simulation (see section \ref{sec:full-scale-sim}. The problem instance is the same as described in the caption of figure \ref{fig:route-map}.}
\label{fig:route-map-new}
\end{figure}

 The truck loop algorithm provides a simple way to break up the enormous logistical routing problem (section \ref{subsec:full-aisin-vrp}) into smaller problems relating to one truck at a time. The output of the truck loop algorithm is a set of routes  $\xi$ as in equation \eqref{eq:truck-loop-output}. These routes were computed with methods that ignored most of the inter-truck interaction and also treated boxes as a continuous ``box soup''. How can we apply the routes $\xi$ in a more realistic fashion given those simplifications?
 
 We chose to interpret the output $\xi$ as ``suggested routes'' and to attempt to use them in a \emph{full-scale} supply chain simulation. The simulation we built does not make any of the simplifications described in section \ref{sec:Single-Truck-PUBO} or \ref{sec:Supply-Chain-Workflow}. Boxes are treated as real boxes with identity. Every box is individually tracked, and every box has given rank-2 or rank-3 requirements. In other words, boxes have required paths as in equation \eqref{eq:box-path}, but those paths have either length 2 or length 3. Boxes also have specific individual volumes. All of this detail is in the supply chain data set of Aisin Corp.
 
 Trucks follow along the routes they were assigned by the truck loop algorithm and they pick up as many boxes as is sensible for their route. The method of pickup selection is very similar to that described for \code{estimate\_demand} in section \ref{sec:Supply-Chain-Workflow}. We briefly describe it here for clarity. Suppose that a truck follows the route $z_1, z_2, z_3, z_4$. When the truck starts (at $z_1$), it first tries to pick up demand destined for $z_2$. We thus look at all boxes currently at $z_1$ with their next stop being $z_2$. As many such boxes (with no bias for which ones) are picked up as can fit on the truck. Then truck truck asks if there is space for boxes at $z_1$ that are destined for $z_3$ that still fit on the truck. The same is done for $z_4$ if there is remaining space. At this point, the truck drives to $z_2$. \emph{During this time, other trucks will be driving and will alter the material at various nodes--we simulated all of this detail.} When the truck arrives at $z_2$, we first drop off all boxes that need to go to $z_2$ as their next stop. We then repeat the prior procedure, asking if there are boxes at $z_2$ that need to go to $z_3$ and so on. Our simulation also takes into account exact details of driving windows, ensuring that no driving goes outside of these windows.
 
 This detailed simulation converts a list of suggested routes for trucks into a precise history of what every truck in a supply chain does and the path of every box in the supply chain. Such simulation is an important tool because it allows algorithms like our truck loop algorithm to make various simplifications while still getting a final result that is commercially useful.
 
 One more utility of the exact simulation, is that it allows for making final corrections to routes and knowing for a fact that those corrections improve performance. For our purposes, we used a very simple heuristic to modify routes: after running a simulation, we checked to see if there were trucks which end their routes on a sequence of steps carrying no demand whatsoever. We clipped these parts of the routes away, and replaced them with a route driving back and forth between whatever two nodes had the highest unsatisfied demand during that simulation run. We then performed the simulation again, and kept it if the overall satisfied demand was superior. We performed this step only 5 times for each overall run, and found only modest improvements.
 
 For both the simulated annealing routes and the D-Wave Hybrid routes, we kept only the first 61 trucks. We found that these trucks satisfied $96.45\%$ of the overall Aisin Corp. demand when using simulated annealing and $99.39\%$ with the D-Wave Hybrid solver. We make no claim that these results would compete with dedicated classical optimization tools commonly used in operations research. However, these results illustrate the viability of breaking problems into small pieces, running those pieces on solvers that can only run smaller problem sizes (like near-term quantum algorithms and quantum annealers), and restoring a realistic solution through simulation.
 
 The truck routing found through our methods using the D-Wave hybrid solver
 followed by the full-scale simulation with 61 trucks is depicted in figure \ref{fig:route-map-new} whereas the routes devised by Aisin Corporation
logistics experts are shown in figure \ref{fig:route-map}. The truck routes
that we found have a substantially greater degree of connectivity than
those currently used in practice.

\subsection{Comparison with Current Supply Chain Performance}
\label{sec:expert-comparison}

 \begin{figure}
\centering
\includegraphics[width=.45\textwidth]{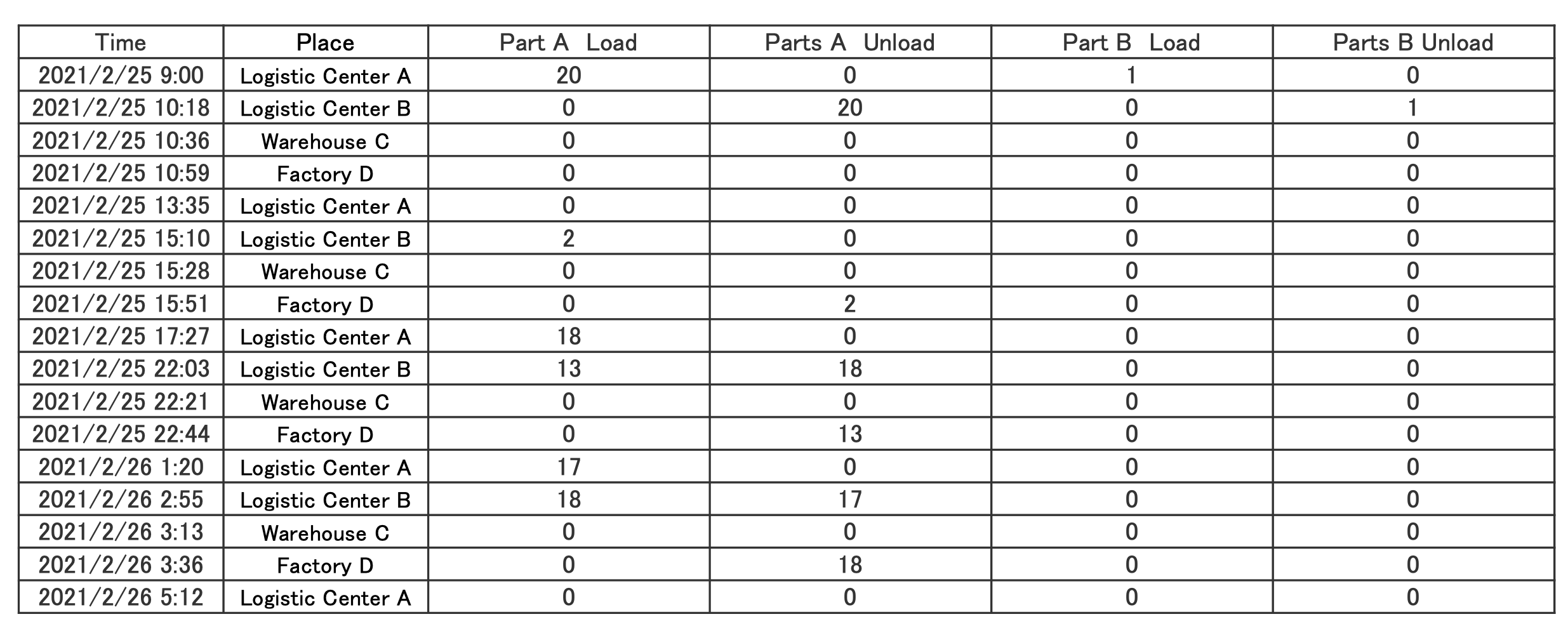}
\caption{Example of a detailed route found for a truck after
D-Wave hybrid execution followed by full-scale simulation. This table
only includes loading and unloading instructions for two of approximately $15,000$ parts, all of which are accounted for in our workflow.}
\label{fig:route-example}
\end{figure}

Aisin Corporation currently routes trucks in their supply chain
by careful analysis by logistics experts. In this subsection,
we briefly compare the current routes used by Aisin Corporation with
the routes determined by the execution of our workflow. A major finding
is that our routes appear to outperform current routing by a number of
metrics.

The connectivity graph currently used by Aisin Corporation 
(that is, the pairs of nodes that trucks currently drive between)
is shown in figure \ref{fig:route-map}. Our execution on the
D-Wave hybrid solver, followed by full-scale simulation with 61 trucks
as described in sections \ref{sec:execution} and \ref{sec:full-scale-sim},
yielded the substantially denser connectivity graph shown in figure \ref{fig:route-map-new}. The routes currently used involve 142 trucks
whereas our methods use only 61 trucks, implying a large reduction in cost. 

These are exciting results, but
we caution that our simulations are not complete in all regards. There
may be constraints that logistics experts must account for that we
have not been able to include in the simulation of section \ref{sec:full-scale-sim}, so the overall reduction in truck number may not
be as dramatic as it appears. Nonetheless, using our routes as a starting point
and making small adjustments is expected to yield a meaningful cost reduction.

A less obvious benefit of the workflow we describe is that it yields very
precise instructions for each truck's route and delivery requirements.
An example of such a routing is shown in figure \ref{fig:route-example}.
(This figure only shows the requirements for two parts, referred to as part A
and part B.) Because there are approximately 15,000 parts, such a precise
prescription is extremely useful. Aisin Corporation currently relies on
human experience and intuition to make precise determinations for
part pickup requirements, while our approach removes this pain point. We emphasize, however, that specific pickup and dropoff instructions are
determined by the full scale simulation implementation, and are not directly
determined by the PUBO optimizer.
 
 \section{Conclusions}

Quantum algorithms are known to provide computational benefits
over classical computing for specific tasks. This specificity
implies that, even when quantum hardware matures, quantum computing
will most likely be used as a component in hybrid classical-quantum
workflows when dealing with complex problems of practical significance.

The methodology of this work is an example of such a hybrid workflow.
Even for the enormously complex task of routing trucks in a realistic
supply chain, we were able to find a way to construct small problem instances
as subroutines that can be solved with NISQ quantum algorithms. For our
study of a commercial supply chain, the
smaller problems are binary optimization problems with $\sim 2500$ binary variables, an appropriate scale for near-term
quantum annealing and somewhat near-term circuit model quantum computing. When
running our workflow with solvers like simulated annealing and the D-Wave
hybrid solver, proxies used in lieu of more mature quantum hardware,
we found viable solutions for the full supply chain.

We do not claim that our method furnishes a provable performance advantage
over classical algorithms because NISQ quantum optimization techniques like
the quantum annealing and QAOA are heuristics without relevant 
proven guarantees. However, the individual truck routing binary optimization
problem is itself NP-hard, and can thus be a performance bottleneck. It's
sensible to apply quantum algorithms to these bottlenecks, and this
approach will pay off if NISQ quantum optimization algorithms that
outperform classical computing are discovered.

Beyond vehicle routing, we believe that our approach is viable
for a wide range of practically important optimization problems.
Many problems can be approximately solved through heuristics 
that decompose a large
problem into solving a large number of smaller problems, and such
smaller problems can often be reduced to binary optimization.
As long as the number of variables for those small problems 
is appropriate for NISQ quantum algorithms, such a hybrid 
approach can be used as a way to explore
the performance of quantum optimization algorithms in the use 
case without resorting to miniaturizing the actual problem.
By building such workflows, industries can deploy optimization
algorithms that can employ quantum algorithms as subroutines
and, in this way explore the potential value of quantum computing
in a variety of use-cases.

 \section*{Disclosure of Financial Interests}
The authors declare no competing non-financial interests.  The authors do disclose the following financial interests.
S.W., F.S., and R.C. were all employed by QC Ware Corp. during their scientific and writing contributions to this manuscript. R.C. is a co-founder of QC Ware Corp. S.W., F.S., and R.C. hold QC Ware stock or stock options. T.I. and K.K. were employed by Aisin Group throughout their respective scientific and writing contributions. Aisin Group is a customer of QC Ware; Aisin Group provided funding for the research presented in this manuscript. 

 \bibliographystyle{unsrt}
 \bibliography{bibliography.bib}

\begin{thebibliography}{10}

\bibitem{shor1999polynomial}
Peter~W Shor.
\newblock Polynomial-time algorithms for prime factorization and discrete
  logarithms on a quantum computer.
\newblock {\em SIAM review}, 41(2):303--332, 1999.

\bibitem{feynman2018simulating}
Richard~P Feynman.
\newblock Simulating physics with computers.
\newblock In {\em Feynman and computation}, pages 133--153. CRC Press, 2018.

\bibitem{lloyd1996universal}
Seth Lloyd.
\newblock Universal quantum simulators.
\newblock {\em Science}, pages 1073--1078, 1996.

\bibitem{arute2019quantum}
Frank Arute, Kunal Arya, Ryan Babbush, Dave Bacon, Joseph~C Bardin, Rami
  Barends, Rupak Biswas, Sergio Boixo, Fernando~GSL Brandao, David~A Buell,
  et~al.
\newblock Quantum supremacy using a programmable superconducting processor.
\newblock {\em Nature}, 574(7779):505--510, 2019.

\bibitem{farhi2014quantum}
Edward Farhi, Jeffrey Goldstone, and Sam Gutmann.
\newblock A quantum approximate optimization algorithm.
\newblock {\em arXiv preprint arXiv:1411.4028}, 2014.

\bibitem{kadowaki1998quantum}
Tadashi Kadowaki and Hidetoshi Nishimori.
\newblock Quantum annealing in the transverse ising model.
\newblock {\em Physical Review E}, 58(5):5355, 1998.

\bibitem{farhi2000quantum}
Edward Farhi, Jeffrey Goldstone, Sam Gutmann, and Michael Sipser.
\newblock Quantum computation by adiabatic evolution.
\newblock {\em arXiv preprint quant-ph/0001106}, 2000.

\bibitem{peruzzo2014variational}
Alberto Peruzzo, Jarrod McClean, Peter Shadbolt, Man-Hong Yung, Xiao-Qi Zhou,
  Peter~J Love, Al{\'a}n Aspuru-Guzik, and Jeremy~L O’brien.
\newblock A variational eigenvalue solver on a photonic quantum processor.
\newblock {\em Nature communications}, 5(1):1--7, 2014.

\bibitem{preskill2018quantum}
John Preskill.
\newblock Quantum computing in the nisq era and beyond.
\newblock {\em Quantum}, 2:79, 2018.

\bibitem{johnson2011quantum}
Mark~W Johnson, Mohammad~HS Amin, Suzanne Gildert, Trevor Lanting, Firas Hamze,
  Neil Dickson, Richard Harris, Andrew~J Berkley, Jan Johansson, Paul Bunyk,
  et~al.
\newblock Quantum annealing with manufactured spins.
\newblock {\em Nature}, 473(7346):194--198, 2011.

\bibitem{dantzig1959truck}
George~B Dantzig and John~H Ramser.
\newblock The truck dispatching problem.
\newblock {\em Management science}, 6(1):80--91, 1959.

\bibitem{toth2002vehicle}
Paolo Toth and Daniele Vigo.
\newblock {\em The vehicle routing problem}.
\newblock SIAM, 2002.

\bibitem{karp1972reducibility}
Richard~M. Karp.
\newblock Reducibility among combinatorial problems.
\newblock {\em Complexity of Computer Computations: Proceedings of a symposium
  on the Complexity of Computer Computations}, pages 85--103, 1972.

\bibitem{lucas2014ising}
Andrew Lucas.
\newblock Ising formulations of many np problems.
\newblock {\em Frontiers in physics}, 2:5, 2014.

\bibitem{pincus1970letter}
Martin Pincus.
\newblock Letter to the editor—a monte carlo method for the approximate
  solution of certain types of constrained optimization problems.
\newblock {\em Operations research}, 18(6):1225--1228, 1970.

\bibitem{van1987simulated}
PJM Van~Laarhoven and EHL Aarts.
\newblock Simulated annealing: theory and applications. dordrecht: D.
\newblock {\em Reidel Pub. Comp., Netherlands}, 1987.

\bibitem{aharonov2004adiabatic}
D.~Aharonov, W.~van Dam, J.~Kempe, Z.~Landau, S.~Lloyd, and O.~Regev.
\newblock Adiabatic quantum computation is equivalent to standard quantum
  computation.
\newblock pages 42--51, 2004.

\bibitem{santoro2006optimization}
Giuseppe~E Santoro and Erio Tosatti.
\newblock Optimization using quantum mechanics: quantum annealing through
  adiabatic evolution.
\newblock {\em Journal of Physics A: Mathematical and General}, 39(36):R393,
  2006.

\bibitem{denchev2016what}
Vasil~S. Denchev, Sergio Boixo, Sergei~V. Isakov, Nan Ding, Ryan Babbush, Vadim
  Smelyanskiy, John Martinis, and Hartmut Neven.
\newblock What is the computational value of finite-range tunneling?
\newblock {\em Phys. Rev. X}, 6:031015, Aug 2016.

\bibitem{dwave2017advantage}
Catherine McGeoch and Paul Farré.
\newblock Advantage processing overview, a technical report.
\newblock {\em White paper}, 2017.

\bibitem{adame2018inhomogeneous}
Juan~I. Adame and Peter~L. McMahon.
\newblock Inhomogeneous driving in quantum annealers can result in
  orders-of-magnitude improvements in performance.
\newblock 2018.

\bibitem{dwave2021hybrid}
Hybrid solver for discrete quadratic models.
\newblock {\em White paper}, 2021.

\bibitem{qubovert}
{Iosue, Joseph T.}
\newblock Qubovert.
\newblock v1.2.4, https://pypi.org/project/qubovert/.

\end{thebibliography}

 \end{document}